\documentclass[5p,twocolumn]{elsarticle}

\usepackage{graphicx}
\usepackage{subfigure}
\usepackage{longtable}
\usepackage{graphicx,wasysym}
\usepackage{longtable}
\usepackage{natbib}
\bibpunct{(}{)}{;}{a}{}{,}
\usepackage{amssymb}

\usepackage[usenames]{color}

\usepackage[figuresright]{rotating}
\usepackage{lscape}
\def\hess{H.E.S.S.\ }

\def\msol{M_{\odot}}
\def\gr{$\gamma$-ray\ }
\def\grs{$\gamma$-rays\ }

\def\wco{W_\mathrm{CO}}
\def\whi{W_\mathrm{HI}}

\journal{JHEAP}

\begin{document}
\begin{frontmatter}
  
\title{Estimating Galactic gas content using different tracers: \\
Compatibility of results, dark gas, and  
unidentified TeV sources}
\author[ieec]{Giovanna Pedaletti\corref{cor1}}
\author[ieec]{Emma de O\~{n}a Wilhelmi}
\author[ieec,icrea]{Diego F. Torres}
\author[mpik,jhi]{Giovanni Natale}
\address[ieec]{Institute of Space Sciences (IEEC-CSIC),
              Campus UAB,  Torre C5, 2a planta,
              08193 Barcelona, Spain} 
\address[icrea]{Instituci\'o Catalana de Recerca i Estudis Avan\c{c}ats (ICREA)}
\address[mpik]{Max-Planck-Institut f\"ur Kernphysik, P.O. Box 103980, D 69029
Heidelberg, Germany}
\address[jhi]{Jeremiah Horrocks
Institute, University of Central Lancashire, Preston, PR1 2HE, UK}
\cortext[cor1]{now at DESY, Platanenallee 6, 15738 Zeuthen, Germany}
\begin{abstract}
A large fraction of Galactic very-high energy (VHE; E$\gtrsim$100 GeV) \gr sources is cataloged as unidentified. In this work we explore 
the possibility that these unidentified sources are located in ambients particularly rich in material content unaccounted by traditional tracers. In a scenario where the VHE emission is due to the interaction of the accelerated particles with a target mass, a large mass of untraced material could be substantially contributing to the VHE emission from these regions. Here, we use three tracers for the commonly explored components: intensity of the $^\textrm{12}$CO(1$\rightarrow$0) line to trace the molecular material, HI hyperfine transition at 21cm to trace atomic hydrogen, and dust emission to trace the total hydrogen content. We show that the estimates of material content from these three tracers are compatible if the uncertainty on the respective conversion factors is taken into account. No additional gas component is found in these regions. However, a simple mass estimation from the $^\textrm{12}$CO(1$\rightarrow$0) line intensity might underestimate the total mass component in some locations.

\end{abstract}


\end{frontmatter}

\section{Introduction}

Up to date there are 84 very-high energy (VHE; E$\gtrsim$100 GeV) sources with Galactic latitude $|b|<5^\circ$. Most of these sources have been discovered through the \hess scan of the Galactic region \citep[HGPS, ][]{hgps_icrc13}. 
A compilation of source properties detected by different experiments can be found at the web resource TeVCat\footnote{at http://tevcat.uchicago.edu/}.
While in most cases the identification of the object responsible for VHE emission is possible thanks to a multiwavelength investigation of the region, at least 25 detections have no firm counterpart. From here onwards we will refer the latter as unidentified VHE sources (UNID).

The molecular content is usually estimated through the intensity of the $^\textrm{12}$CO(1$\rightarrow$0) line at 115 GHz \citep[2.6 mm, see e.g.,][]{dame_co}. However, if the gas is diluted, this method is shown to fail \citep{planckdark}. The gas that cannot be traced with this method is usually referred to as \textit{``dark''} gas, or \textit{``CO-faint''} gas \citep{grenierdark}. Existence of such gas co-located with the UNID TeV sources may relax conditions on the accelerator power that is putatively injecting cosmic-rays, the interaction of which would generate the TeV photons. VHE sources are also expected to be associated with regions of high stellar activity, that are generally regions with high material content and high emission from the dust component. 
Then, in order to better trace the total material content of a region, dust emission and an HI content estimation should also be used. The HI content can be traced through the intensity of the 21cm line due to the hyperfine transition at the ground level. Dust emission can be traced through its thermal emission peaking at far infrared wavelenghts ($\nu_\mathrm{dust}\sim 200$ GHz). In the following, we will use all three methods to detail the environment on line-of-sight of UNID sources.

We frame this study in a larger scenario by analyzing with a similar method the compatiblity of results for gas content estimations along every line of sight in the Galaxy.

The procedure for the calculation of material content is explained in Section \ref{sec:estimation}. In Section \ref{sec:define_and_gal} we define the parameterization we use and explore the material content relations in the Galactic plane. Details of the selected UNID sources and of the related material content are given in Section \ref{sec:unid}. Control populations are explored in Section \ref{sec:control} and conclusions are given afterwards.

\section{Material content estimation}\label{sec:estimation}
\begin{figure*}
\begin{center}
\includegraphics[width= 1. \linewidth]{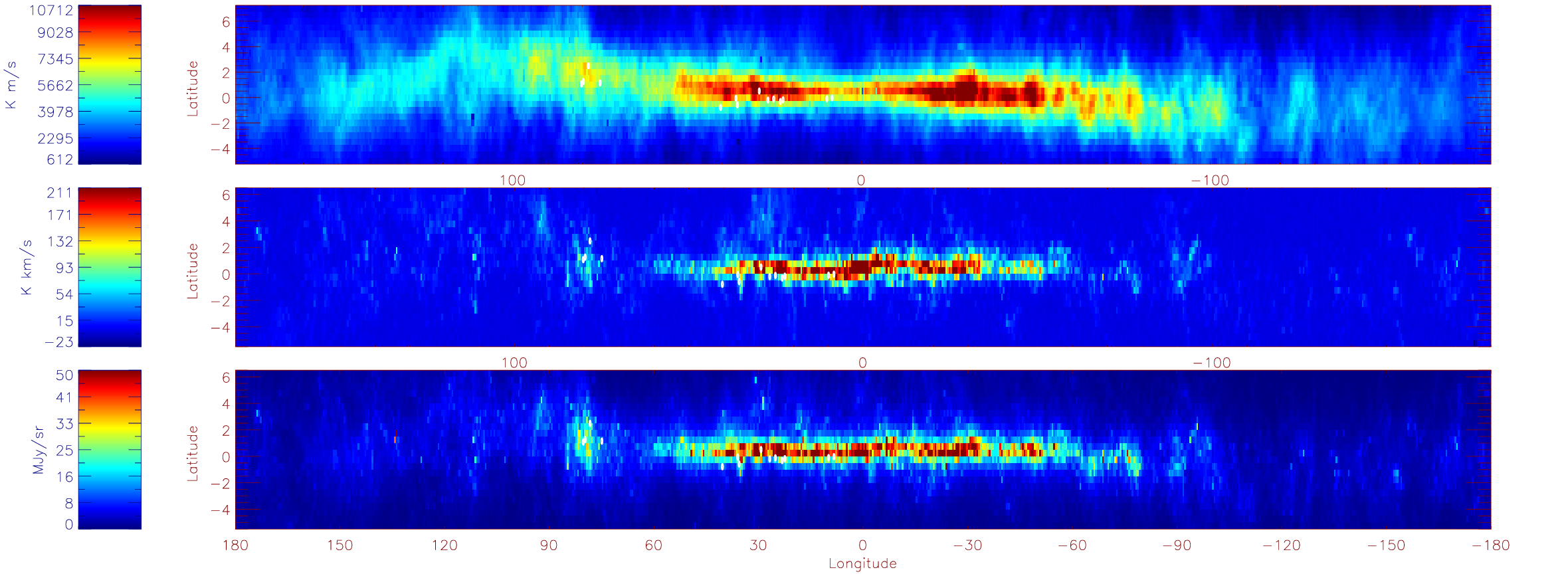}
\caption{Maps in Galactic coordinates of the tracers used in this paper. From top to bottom: \textbf{top}, HI 21 cm line intensity integrated along the line of sight, from the LAB survey \citep{lab_hi}; \textbf{middle}, CO line intensity from the Planck repository, ``Type2'' \citep{planckco}; \textbf{bottom}, dust emission intensity at 353 GHz from the Planck repository \citep{planckdust}. All the maps have been scaled so that the minimum and maximum of the scale are within 99\% percent of the mean pixel value, enhancing the more diffuse region.}
 \label{fig:maps}
\end{center}
\end{figure*}

The gas surface density at a given projected position in the sky can be expressed as
\begin{equation}\label{eq:mass}
 \Sigma=\mu m_\mathrm{H} N_\mathrm{H},
\end{equation}
where $N_\mathrm{H}$ is the column density, $\mu=1.4$ is the mean weight per H atom, $m_\mathrm{H}$ is the mass of the H nucleon.

The dust content can be used as tracer of the total mass (hydrogen in any form); i.e., Eq. \ref{eq:mass} is used to derive the total content $\Sigma_\mathrm{dust}$, the surface brightness estimated from dust emission, once $N_{\rm{H}}$, column density of the hydrogen, is calculated from the thermal emission of dust \citep[see, e.g.][]{roy_orion_dust} as follows:
\begin{equation}\label{eq:massdust}
 N_\mathrm{H}= \sigma_{\nu_0}^{-1} \displaystyle{\sum\limits_\mathrm{px}} \tau_\nu \left( \frac{\nu}{\nu_0}\right)^{-\beta}
\end{equation}
where $\sigma_{\nu_0}$ is the dust emissivity at the reference frequency $\nu_0$, which contains the knowledge on the emissivity of the dust grains and their composition, $\tau_\nu$ is the dust optical depth at frequency $\nu$, and $\beta$ is the dust spectral index.
The {\it pixel} (px) refers to the unit division of the map that will be used to obtain the value of $\tau_\nu$, in this case pixels of $0.5^\circ$, see below.
The suffix of $\Sigma_\mathrm{dust}$ refers to the method of estimation, but the value will represent the total gas component.
The intensity of the dust emission can be parametrized as 
\begin{equation} \label{eq:inu_dust}
I_\nu = \tau_\nu B_\nu(T),
\end{equation}
where $B_\nu(T)$ is the Planck function for dust equilibrium temperature $T$. 
The Planck legacy archive\footnote{http://pla.esac.esa.int/pla/aio/planckProducts.html} provides dust parameters as all-sky maps, i.e.,  $\tau$, $\beta$ index and $T$ 
(with degeneracy between the last 2 parameters). $\tau$ is derived keeping the value of $\beta$ fixed. They were obtained by fitting the Planck data at 353, 545 and 857 GHz together with the IRAS (IRIS) 100 micron data \citep[see][]{planckdust}. One can use all-sky maps for $\tau$ to calculate the gas surface density. In the following we use the emission at $\nu_0=$353 GHz. The all-sky maps released in the online archive refer to this frequency $\nu_0$. In Fig. \ref{fig:maps} we show the intensity map of $I$(353GHz) from the Planck legacy archive, for a cut of the Galactic region ($|b|<6^\circ$).

The Planck satellite legacy archive also provides $\wco$ in galactic coordinates, where $\wco$ is the intensity of the CO line (1$\rightarrow$0) at 115 GHz (2.6 mm). We choose here the Planck maps with a higher signal to noise ratio, tagged as ``Type2'' in the archive \citep{planckco}.
From $\wco$ one can estimate the amount of molecular material as follows: 
\begin{equation}\label{eq:massco}
 N_\mathrm{H2}=X_\mathrm{CO} \displaystyle{\sum\limits_\mathrm{px}} \wco.
\end{equation}
Indeed, the definition of the conversion factor $X_\mathrm{CO}=N_\mathrm{H2}/\wco$ comes from the assumption that the content of CO is proportional to the molecular content of the region \citep{xcoreview}. It must be noted that the linearity of Eq. \ref{eq:massco} is assured for a large range of column densities. The linearity breaks down for region of very high or very low densities. This translates into the necessity of adopting different values of $X_\mathrm{CO}$ for different environments. The possible values will be described in Section \ref{sub:macxcoconstants}. Eq. \ref{eq:mass} is then used to translate the column density, $N_\mathrm{H2}$ (H2 indicates only the molecular component), to the total molecular gas content $\Sigma_\mathrm{CO}$, the surface brightness estimated from CO content. Again, the suffix CO refers only to the tracer. {\it Pixel} (px) refers to the unit division of the map in question, that was also binned to px=$0.5^\circ$, as before. 

The last tracer used allows the estimation of the atomic component: the intensity of the 21cm HI line.
This quantity is provided by the Leiden Bonn Argentine (LAB) survey, which provides mapping of the intensity of the 21cm HI line in cubes of galactic latitude, longitude, and radial velocity \citep{lab_hi}. The radial velocity in local standard of rest of the map would allow the estimation of distances after the assumption of a Galactic rotation model. The information for the dust map is two-dimensional, giving the intensity of thermal emission for each location in the projected sky. Therefore, in the following, we simply integrate the $\whi$ information along the line of sight.
One can use the values from LAB survey, convert the velocity-integrated intensity into $N_{\rm{HI}}$ and then use Eq. \ref{eq:mass} to calculate the atomic content, i.e 
\begin{equation}\label{eq:masshi}
N_{\rm{HI}}=X_\mathrm{HI} \displaystyle{\sum\limits_\mathrm{px}}\whi. 
\end{equation}
The sum of $\displaystyle{\sum\limits_\mathrm{px}}\whi$ is calculated from the brightness temperature given in the dataset and the binning in velocity of the data cube. The conversion factor in this case is $X_\mathrm{HI}=1.82\times 10^{18} \textrm{cm}^{-2} (\textrm{K km/s})^{-1}$ \citep[see, e.g., Eq. 1 of][]{fukuithick}, valid in the optically thin approximation. However, the mass of the HI component can be underestimated if the optically thin approximation for the 21 cm line is no longer valid (at large column densities). As described by \citet{fukuithick}, this can lead to the mass estimate to be off by a factor of $X_\mathrm{HI,thick}\sim2X_\mathrm{HI}$. 

Fig. \ref{fig:maps} shows the maps we use, for a cut of the Galactic region ($|b|<6^\circ$), pixelized with our search radius at the different frequencies. The pixels are of $0.5^\circ\times0.5^\circ$. Using the healpix package\footnote{http://healpix.sourceforge.net/}, the dust and $\wco$ maps have been repixelized to match the resolution of the LAB survey maps (1px=0.5$^\circ$). This is slightly larger than the original pixelization of the all-sky map of optical depth \citep[5', ][]{planckdust}. This extension was chosen to have a consistent pixelization in all maps, but also to have an extension comparable to a circle of $0.22^\circ$ in radius, that was chose as the discovery radius in the HGPS \citep{surveyhess}.


\subsection{Conversion parameters for gas surface density estimates}\label{sub:macxcoconstants}

The emissivity of the dust ($\sigma_{\nu_0}$) is a difficult parameter to estimate, and depends on the type of dust considered and on the density of the region. A reliable calculation of the hydrogen column density from Eq. \ref{eq:massdust} would require the knowledge of $\sigma_{\nu_0}$ tied to the value of the $\beta$ index, i.e. both $N_\mathrm{H}$ and $\sigma_{\nu_0}$ need to be derived with the same assumption on the $\beta$ index. If the dust emission SED is fit by
leaving $\beta$ as a free parameter and the dust mass is derived assuming a dust cross section for a different value of $\beta$, then the mass estimate can be wrong by a factor 2-3. Instead, if the SED is fit using correct combinations of $\beta$ and dust cross section,
the estimated dust masses can vary by only 10-20\% independently from the
specific value of $\beta$ used \citep{bianchi_beta}. All the dust related quantites investigated here assume $\beta=1.8$, so that our calculation of gas surface density is self consistent. 
\citet{roy_orion_dust}  show that the dust opacity $\sigma_{1200 GHz}$ increases with column density ($\sigma_{1200 GHz} \propto N_{\rm H}^{0.28 \pm 0.01 \pm 0.03}$). 
In \citet{ismplanck} the best estimate of the dust opacity is $\sigma_{\rm e}= (0.92 \pm 0.05) \times 10^{-25}$  cm$^2$ H$^{-1}$ at 250 $\mu$m (1200GHz), for the ISM, while in the Orion molecular cloud the opacities can be larger by a factor of $\sim 2-4$ \citep[Herschel data from][]{roy_orion_dust}.
A study of the impact of dust grain composition in a laboratory setting is given in \cite{macstudy}, where values as high as $\kappa_{850\mu m}\simeq0.8 \textrm{ m}^2/\textrm{kg}$ are shown for the low temperature relevant for dust. This would correspond to $\sigma(353\mathrm{GHz})=\mu m_{\rm H} \kappa_\nu r =2 \times 10^{-25}$ cm$^2$ H$^{-1}$, with $r=0.01$ being the dust-to-gas mass ratio. This is a very extreme value, relevant for only one of the synthesised analogues of interstellar amorphous silicate grain samples, and we will not use it here.

Therefore, parameterizing $\sigma_{\rm e}(\nu)/\sigma_{\rm e} (\nu_0) = (\nu/\nu_0)^{\beta}$ and considering the maximum value for the opacity as $\sigma_{\rm e}(1200\mathrm{GHz})=3.7 \times 10^{-25}$ cm$^2$ H$^{-1}$, we obtain
 $\sigma_{high}(353\mathrm{GHz})=4.1 \times 10^{-26}$ cm$^2$ H$^{-1}$ for $\beta=1.8$ compatible with \citet{roy_orion_dust}. Similarly, considering as a lower value the ISM estimate, this translate in a value of $\sigma_{low}(353\mathrm{GHz})=1 \times 10^{-26}$ cm$^2$ H$^{-1}$.

Regarding the molecular content tracer, we will consider the range $X_\mathrm{CO} = [0.5\ldots4.8] \times 10^{20} \textrm{cm}^{-2} (\textrm{K km/s})^{-1}$ \citep{xcoreview}. We note that \citet{xcoreview} cites values as low as $X_\mathrm{CO}= 0.5 \times 10^{20} \textrm{cm}^{-2} (\textrm{K km/s})^{-1} $ only for particular regions in the central part of the Galaxy. However, most of the UNID sources that we want to study here are located at a galactocentric distance larger than 500 pc with the possible exception of source \#10 (see Table \ref{tbl:unid} below). The recommended standard Galactic value is $X_\mathrm{CO}= 2 \times 10^{20} \textrm{cm}^{-2} (\textrm{K km/s})^{-1} $ \citep{xcoreview}.

Thus, to summarize we consider the following intervals:
\begin{eqnarray}\label{eq:limits}
\sigma (353\mathrm{GHz})&=& \{ 1. - 4.1 \} \times 10^{-26} \mathrm{cm}^2 \ \mathrm{H}^{-1} \nonumber  \\
X_\mathrm{CO} &=& \{0.5 - 4.8\}  \times 10^{20} \textrm{cm}^{-2} \ (\textrm{K km/s})^{-1}  
\end{eqnarray}

For ease of interpretation however, we will also refer to the average value of these parameters as:
\begin{eqnarray}\label{eq:averages}
\sigma (353\mathrm{GHz})&=& 2.5 \times 10^{-26} \mathrm{cm}^2 \ \mathrm{H}^{-1} \nonumber  \\
X_\mathrm{CO} &=& 2  \times 10^{20} \textrm{cm}^{-2} \ (\textrm{K km/s})^{-1}  
\end{eqnarray}
\section{Relations among the gas surface density estimates}\label{sec:define_and_gal}

Requesting that different estimates trace the same total content
\begin{equation}
\Sigma_\mathrm{dust} = \eta \Sigma_\mathrm{CO},
\end{equation}
one can define the quantity $R_\mathrm{dust/CO}$:
\begin{equation}\label{eq:ratiounsig}
  R_\mathrm{dust/CO}\equiv\eta X_\mathrm{CO} \sigma_{\nu_0}= \frac{\displaystyle{\sum\limits_{px}}\tau_\nu (\nu/\nu_0)^{-\beta}}{\displaystyle{\sum\limits_{px}} \wco},
\end{equation}
that we will call dust-to-CO ratio, where $\eta$ should be $\sim 1$ in the case that molecular material
is dominant (i.e., $N_\mathrm{HI} \ll N_\mathrm{H2}$). 
$R_\mathrm{dust/CO}$ is defined as such from the ratio of the gas surface densities, but separating the observables (dust emission and CO line intensity) from the conversion parameters. The values of the observables are derived from the tracer maps as detailed above. The range of possible values for the ratio $R_\mathrm{dust/CO}$ is 
\begin{equation}\label{eq:bandratio}
R_\mathrm{dust/CO} = \{0.1\ldots4\},
\end{equation}
when considering the interval ranges for the dust emissivity and $X_\mathrm{CO}$ conversion factor derived in \ref{sub:macxcoconstants}, see Eq. \ref{eq:limits}, and normalizing to the average values given in Eq. \ref{eq:averages}. Thus the average value of the dust-to-CO ratio parameter will be $R_\mathrm{dust/CO} =1$.
Extremes of this range of $R_\mathrm{dust/CO}$ pinpoint mismatches in the mass estimations. 
When $R_\mathrm{dust/CO} \geq 4$, there is an excess of dust emission with respect to what can be expected given the $\wco$ estimated at the same location. In other words, even using a larger dust emissivity, the estimate of gas surface density coming from dust will still exceed the estimate from $\wco$. Or, conversely, there is less $\wco$ than expected from the dust emission. 
On the other hand, if $R_\mathrm{dust/CO} \leq 0.1$, there is a stronger intensity $\wco$ than what can be expected given the dust emission at the same position. The $\Sigma_\mathrm{CO}$ traces best the molecular component, while $\Sigma_\mathrm{dust}$ traces all the material. Assuming $\eta=constant$ is equivalent to assume that the dust and the CO line intensity can both trace the entire material content. It seems sensible to expect that the $R_\mathrm{dust/CO}$ at each line of sight in the Galaxy be contained in the band given by Eq. \ref{eq:bandratio}. 

\begin{figure}
\begin{center}
\includegraphics[width=0.95 \linewidth]{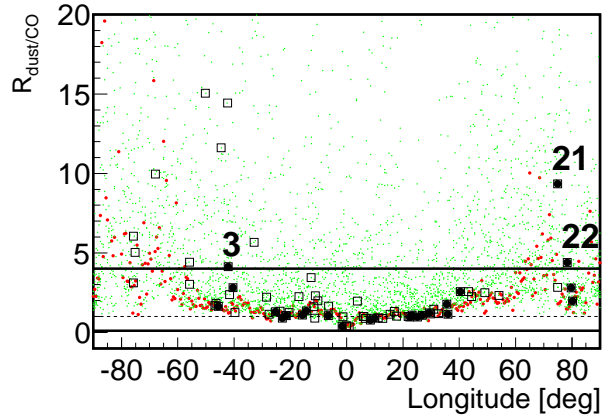}
\caption{$R_\mathrm{dust/CO}$ versus galactic longitude. The horizontal solid lines delimit the accepted region for the value of $R_\mathrm{dust/CO}$, with its average value represented by the dashed thin line. Red and green points are calculated for each $0.5^\circ$ pixel in the maps with $|l|<90^\circ$ and $|b|<0.5^\circ$ or $0.5^\circ<|b|<3^\circ$, respectively.  Black solid points are for the sample of UNID objects. Numbers of the objects outside the range refer to the list in Table \ref{tbl:unid} and belong to regions outside the galactic plane ($|b|>0.5$). For comparison we show also the larger Galactic VHE sample with open black squares, see Sec. \ref{sec:unid}.}
 \label{fig:unidratiosigma}
\end{center}
\end{figure}

We have calculated the value of $R_\mathrm{dust/CO}$ in each pixel of the maps with $|l|<90^\circ$ and $|b|<3^\circ$. The values can be seen in Fig. \ref{fig:unidratiosigma} in red and green. The green points in Fig. \ref{fig:unidratiosigma} represent the values of  $R_\mathrm{dust/CO}$ for each line of sight in the Galaxy with $|b|>0.5^\circ$ and red otherwise.  

We can think of the Galactic longitude as a proxy for both Galactocentric distance and column density in the Galactic plane.
It has been shown that the range of $X_\mathrm{CO}$ is actually due to an increase from a minimum value at small galactocentric distances \citep{xcoreview}. 
The V-shaped distribution seen in Fig. \ref{fig:unidratiosigma} might be a reflection of this $X_\mathrm{CO}$ behavior, with lower values of $R_\mathrm{dust/CO}$ at small longitudes.
The dust opacity $\sigma_{\nu}$ has an opposite behavior, i.e., it  increases with column density (naively forcing towards an inverted V-shape, with higher values of $R_\mathrm{dust/CO}$ at small longitudes). However, if one were to exclude the very central region of the Galaxy, there is no dominant trend of $R_\mathrm{dust/CO}$ with Galactic longitude, hence we do not try to refine the band depending on sky position. 

We remind the meaning of $R_\mathrm{dust/CO}$ limiting values: the upper border implies 
a high column density at large galactocentric distance (it could happen in isolated clouds, but certainly it is not the common case); the lower boundary is an even more unprobable case of a low column density at small galactocentric distance.

The limiting values of $R_\mathrm{dust/CO}$ should be regarded as conservative in the case of $\eta=1$.
Nonetheless, many locations are outside the upper bounds defined in Eq. \ref{eq:bandratio}, so to render these bounds not valid outside the inner Galactic plane (see Fig. \ref{fig:unidratiosigma}). It is indeed to be noted also that most of the locations ouside the bounds have either large Galactic longitude or latitude.
This is especially interesting considering the fact such directions in the sky integrate less amount of material along the line of sight, hence the co-location of dust and molecular estimator should be clearer. This follows the fact that the dust profile from the Galactic center falls less sharply than the intensity of the $^\textrm{12}$CO(1$\rightarrow$0) and this effect cannot be mitigated even allowing for differences in conversion parameters.  

The assumption of $\eta=1$ is clearly failing for locations outside the band and indicates that the molecular content is not dominant and we need to consider a further gas component. We therefore consider in the following an obvious candidate for this: the atomic gas component traced through the intensity of the 21cm HI line as described in Sec. \ref{sec:estimation}.

\begin{figure}[!htbp]
\begin{center}
\includegraphics[width=1. \linewidth]{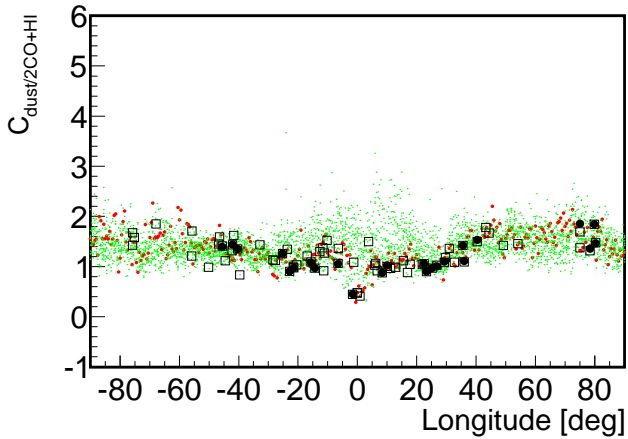}
\caption{Dust estimation excess $C_\mathrm{dust/2CO+HI} $ as a function of Galactic longitude. All points are calculated with $X_\mathrm{CO} = 2.0 \times 10^{20} \textrm{cm}^{-2} (\textrm{K km/s})^{-1}$ and $X_\mathrm{HI,thin}=1.82\times 10^{18} \textrm{cm}^{-2} (\textrm{K km/s})^{-1}$. All points are derived using $\sigma_{low}(353\mathrm{GHz})=1 \times 10^{-26}$ cm$^2$ H$^{-1}$, where $\Sigma_\mathrm{dust}\propto \sigma_\nu^{-1}$. Red and green points are calculated for each $0.5^\circ$ pixel in the maps with $|l|<90^\circ$ and $|b|<0.5^\circ$ or $0.5^\circ<|b|<3^\circ$, respectively. Black solid points refer to the UNID sample. For comparison we show also the larger Galactic VHE sample with open black squares, see Sec. \ref{sec:unid}.}
 \label{fig:ratiovsratio}
\end{center}
\end{figure}
We now define a new parameterization of the three gas surface densities, so to consider all tracers. This estimate does not depend on the common terms, that are $\mu$ and $m_\mathrm{H}$, so that is proportional to $N_\mathrm{H}/(2N_\mathrm{H2}+N_\mathrm{HI})$ \citep[see Eq. 21][]{xcoreview}. 
After simplification, the quantity to be evaluated is:
\begin{eqnarray}
\label{eq:threetracers}
  && \hspace{-1cm} C_\mathrm{dust/2CO+HI} = \frac{N_\mathrm{H}}{2N_\mathrm{H2}+N_\mathrm{HI}}  \nonumber \\
  && =\frac{\sigma_{\nu_0}^{-1}\displaystyle{\sum\limits_{px}}\tau_\nu (\nu/\nu_0)^{-\beta}}{2X_\mathrm{CO}\displaystyle{\sum\limits_{px}} \wco+X_\mathrm{HI}\displaystyle{\sum\limits_{px}} \whi},
\end{eqnarray}
and we will refer to it as dust estimation excess.
We will now fix the dust opacity to $\sigma_{low}(353\mathrm{GHz})=1 \times 10^{-26}$ cm$^2$ H$^{-1}$ (see Eq. \ref{eq:limits}), so to use this value as normalization. In Fig. \ref{fig:ratiovsratio} we show the distribution in longitude of the dust estimation excess derived from Eq. \ref{eq:threetracers}.
We here assume the conversion factor for the molecular component to be $X_\mathrm{CO} = 2. \times 10^{20} \textrm{cm}^{-2} (\textrm{K km/s})^{-1}$, as is the recommended standard Galactic value in \citet{xcoreview}. We then fix the conversion factor for the atomic component to be $X_\mathrm{HI}$, in the optically thin approximation. In this case one can see that, considering the atomic component, most of the UNID source locations are inside an acceptable band of $C_\mathrm{dust/2CO+HI}$, defined as follows. The lower limit of this band is set to $C_\mathrm{dust/2CO+HI}=1$ and corresponds to the minimum value of dust emissivity $\sigma_{low}(353\mathrm{GHz})$, while the upper limit of $C_\mathrm{dust/2CO+HI}\simeq4$, correspond to $\sigma_{high}(353\mathrm{GHz})$ (see Eq. \ref{eq:limits}). There are instances of $C_\mathrm{dust/2CO+HI}<1$. Most cases are in central regions of the inner Galaxy, where a $X_\mathrm{CO} < 2. \times 10^{20} \textrm{cm}^{-2} (\textrm{K km/s})^{-1}$ can be expected (see Section \ref{sub:macxcoconstants}).
Contrary to what is seen in the behaviour of $R_\mathrm{dust/CO}$, also the positions in the sky with large Galactic longitude or latitude are inside the allowed bounds. There is therefore no necessity to invoke a further gas component once one allows for a change in the conversion factor normalization.
\section{Gas content on the direction of UNID TeV sources}\label{sec:unid}
We select all sources that have been tagged as ``UNID'' or ``DARK'' in the TeVCat (Table \ref{tbl:unid}). Source \#23 and \#24, might eventually relate to the same source. However, we stick to the catalog convention of considering them separate, as it is not excluded that the \gr excess might be due to a composition of sources.\footnote{In this work we do not consider the TeVCat entry \textit{``Milagro Diffuse''} even if it is cataloged as unidentified. The very large extension of this diffuse detection would be much larger than the extensions considered here.} We therefore select a sample of 24 sources. We calculate the molecular and total surface density for each of the objects in the way described in Sec. \ref{sec:estimation}.

We calculate the dust-to-CO ratio $R_\mathrm{dust/CO}$ at the position of the UNID TeV sources in Table \ref{tbl:unid} and show that in most cases $R_\mathrm{dust/CO}$ is indeed contained in the band defined in Eq. \ref{eq:bandratio}, see also Fig. \ref{fig:unidratiosigma}. All the objects that have an exceeding $R_\mathrm{dust/CO}$ are located at high Galactic latitudes and especially in the Cygnus region (see Fig. \ref{fig:unidratiosigma} and Table \ref{tbl:unid}), as it was the trend for the general locations of the outer Galactic region explored in Sec. \ref{sec:define_and_gal}. But this is not the case for the inner Galactic region (nor in the case of 2 other samples of interest, where the value of $R_\mathrm{dust/CO}$ is always contained in the defined band, see Section \ref{sec:control}). 

In order to see if the atomic component needs to be taken into account in the locations with a dust-to-CO ratio out of bounds, we now calculate the values of the dust estimation excess parameter $C_\mathrm{dust/2CO+HI}$. As before, we use $\sigma_{low}(353\mathrm{GHz})=1 \times 10^{-26}$ cm$^2$ H$^{-1}$ as normalization. 
We here have assumed the same values of $X_\mathrm{CO}$ and $X_\mathrm{HI}$ as we did for the entire Galactic region explored in Sec. \ref{sec:define_and_gal} (hereafter case A, see Table \ref{tbl:cases}).
In this case one can see that, considering the atomic component, most of the UNID source locations are inside an acceptable band of $1<C_\mathrm{dust/2CO+HI}<4$ (see Sec. \ref{sec:define_and_gal}).  While there is no evidence from this approach of a missing gas component or $C_\mathrm{dust/2CO+HI}>4$, there are instances of $C_\mathrm{dust/2CO+HI}<1$. These locations are in central regions of the inner Galaxy, and most notably, source \#10 in Table \ref{tbl:unid} presents a $C_\mathrm{dust/2CO+HI}=0.45$. 
However, if we assume $X_\mathrm{CO} = 0.5 \times 10^{20} \textrm{cm}^{-2} (\textrm{K km/s})^{-1}$ (herafter case B, see Table \ref{tbl:cases}), a value apppropriate for the Galactic Center region \citep[see][]{xcoreview}, all values would then be inside the limits and specifically $C_\mathrm{dust/2CO+HI}=1.56$ for source number \#10.

To avoid invoking a different dust emissivity from the ISM average in order to reconcile the gas surface densities estimates (hence $C_\mathrm{dust/2CO+HI}\neq1$), other assumptions can reduce the ratio $C_\mathrm{dust/2CO+HI}$. Indeed, the estimate of the CO or HI component can be incremented. With a larger value of $X_\mathrm{CO}$ we can push up $\Sigma_\mathrm{CO}$. The Cygnus region, for example, presents some of the largest values of $C_\mathrm{dust/2CO+HI}$. However this does not seem to be a viable explanation when considering that the average value for the conversion factor in the molecular component in the Cygnus region, derived from Fermi-LAT data at high energies (HE; E$\gtrsim$100 MeV), is $X_\mathrm{CO} = 1.68 \times 10^{20} \textrm{cm}^{-2} (\textrm{K km/s})^{-1}$ \citep{cygnusfermi}. On the other hand, also the mass of the HI component can be underestimated if the optically thin approximation for the 21 cm line is no longer valid (at large column densities). As described by \citet{fukuithick}, 
this can lead to the mass estimate to be off by a factor of $X_\mathrm{HI,thick}\sim2X_\mathrm{HI}$. We take the latter possibility into account (hereafter case C, see Table \ref{tbl:cases}) and 
show the corresponding values of $C_\mathrm{dust/2CO+HI}$ in Table \ref{tbl:unid}. The latter approach would need to be coupled to a $X_\mathrm{CO}$ slightly lower than the Galactic average.

Considering the value of $R_\mathrm{dust/CO}$ and $C_\mathrm{dust/2CO+HI}$, we cannot associate the sample of UNID sources with locations where the canonical estimates of material content are failing. Therefore we cannot link them to dark gas.

The three cases explored here (A, B, C, parameters summarized in Table \ref{tbl:cases}) represent idealized scenarios in which one tries to assign a single value to the parameters at all of the location of the UNID sample simultaneously. 

We have plotted in Fig. \ref{fig:unidratiosigma} and Fig. \ref{fig:ratiovsratio} also the values corresponding to the positions of all of the VHE sources in the same longitude range. It is interesting to see that there are many instances in Fig. \ref{fig:unidratiosigma} of $R_\mathrm{dust/CO}>4$ when considering the identified VHE sources. The sample of identified sources with $R_\mathrm{dust/CO}>4$ consist of a mix of the known VHE emitter classes, with a majority of Pulsar wind nebulae (PWN), all the known binary minus LS5039, and Supernova Remnant (SNR) with shell morphology. First, we have to consider that all of these positions present high Galactic coordinate ($|l|>30^\circ$ and $|b|>0.48^\circ$). In all cases but one, moreover, the detection of these sources did not come from surveying of the Galactic plane or of the Cygnus region, but from long dedicated observations. Thus these are not directly comparable to the rest of the sample here. The exception is HESS J1356-645, that was detected after only 10 hours with \hess observations \citep{1356_hesspaper}. The centroid of the VHE emission from this source is located at with (b=-2.5), hence in a region where the dominance of molecular material on the total gas composition starts to fail (i.e. $\eta\neq1$, see Section \ref{sec:define_and_gal}). When we however calculate the dust estimation excess $C_\mathrm{dust/2CO+HI}$, all of the locations relative to the known VHE sources are inside the bounds, see Fig. \ref{fig:ratiovsratio}. An interesting case is Cassiopeia A that is not shown here, with (b,l)=(111.7,-2.1), that presents $C_\mathrm{dust/2CO+HI}=3.5$. This value is still in the acceptable bounds, but underlines the high emission from dust coming from the environment of CasA, where a high dust emissivity needs to be invoked \citep[see also][]{casa_dust}.

\begin{table*}[!htb]
\centering
\caption{Exemplary cases of conversion factor values}
\begin{tabular}[t]{cccc}
\hline
Case    &  $\sigma(353\mathrm{GHz})$ & $X_\mathrm{CO}$                            & $X_\mathrm{HI}$\\
        &  cm$^2$ H$^{-1}$           & $\textrm{cm}^{-2} (\textrm{K km/s})^{-1}$  & $\textrm{cm}^{-2} (\textrm{K km/s})^{-1}$\\
\hline
\\
A       & $1 \times 10^{-26}$  (low)      &  $2 \times 10^{20}$ (average)                       & $1.82\times 10^{18}$ (opt. thin)\\
B       & $1 \times 10^{-26}$ (low)       &  $0.5 \times 10^{20}$  (low)                    & $1.82\times 10^{18}$ (opt. thin)\\
C       & $1 \times 10^{-26}$  (low)      &  $2 \times 10^{20}$   (average)                     & $3.64\times 10^{18}$ (opt. thick) \\
\hline
\end{tabular}
\label{tbl:cases}
\end{table*}

\begin{table*}[t]
\centering
\caption{Characteristics of the UNID sample. The last column is calculated through Eq. \ref{eq:threetracers} (case A-B-C, see text and Table \ref{tbl:cases} for details)}
\begin{tabular}[t]{cccccc}
\hline
Number & Name    &     glon      &    glat    & $R_\mathrm{dust/CO}$ & $C_\mathrm{dust/2CO+HI}$\\ 
       &         &     (deg)     &    (deg)   &   &  (A-B-C)\\ 
       &         &               &            & \{0.1 ... 4\} &\{1 ... 4\} \\ 
\hline
\\
1&HESSJ1427-608&-45.59&-0.15&1.63&  1.39 -  2.89 -  1.07\\
2&HESSJ1503-582&-40.38& 0.29&2.82&  1.35 -  1.91 -  0.84\\
3&HESSJ1507-622&-42.05&-3.49&4.13&  1.44 -  1.82 -  0.84\\
4&HESSJ1626-490&-25.23& 0.05&1.31&  1.26 -  3.04 -  1.03\\
5&HESSJ1634-472&-22.89& 0.22&0.89&  0.90 -  2.36 -  0.77\\
6&HESSJ1641-463&-21.48& 0.09&1.06&  0.98 -  2.22 -  0.78\\
7&HESSJ1702-420&-15.70&-0.18&1.09&  1.08 -  2.75 -  0.91\\
8&HESSJ1708-410&-14.32&-0.47&1.35&  0.97 -  1.72 -  0.68\\
9&HESSJ1729-345& -6.56&-0.13&1.05&  1.05 -  2.69 -  0.88\\
10&HESSJ1741-302& -1.60& 0.19&0.39&  0.45 -  1.56 -  0.43\\
11&HESSJ1804-216&  8.40&-0.03&0.78&  0.87 -  2.75 -  0.80\\
12&HESSJ1808-204&  9.93&-0.10&0.89&  1.02 -  3.34 -  0.95\\
13&HESSJ1832-093& 22.48&-0.18&0.96&  1.06 -  3.18 -  0.95\\
14&HESSJ1834-087& 23.24&-0.31&1.09&  0.90 -  1.81 -  0.68\\
15&HESSJ1837-069& 25.18&-0.12&0.96&  0.97 -  2.48 -  0.82\\
16&HESSJ1841-055& 26.80&-0.20&1.08&  1.00 -  2.31 -  0.80\\
17&HESSJ1843-033& 29.30& 0.51&1.22&  1.10 -  2.41 -  0.86\\
18&HESSJ1857+026& 35.96&-0.06&1.15&  1.11 -  2.69 -  0.91\\
19&HESSJ1858+020& 35.58&-0.58&1.77&  1.42 -  2.78 -  1.06\\
20&MGROJ1908+06& 40.39&-0.79&2.57&  1.53 -  2.40 -  1.01\\
21&VERJ2016+372& 74.94& 1.15&9.35&  1.85 -  2.11 -  1.01\\
22&VERJ2019+407& 78.33& 2.49&4.39&  1.35 -  1.67 -  0.77\\
23&MGROJ2031+41& 79.72& 1.26&2.79&  1.84 -  3.08 -  1.26\\
24&TeVJ2032+4130& 80.25& 1.07&1.96&  1.46 -  2.67 -  1.05\\
\hline
\end{tabular}
\label{tbl:unid}
\end{table*}
\section{Other samples}\label{sec:control}


\subsubsection{VHE sources in the inner Galaxy}\label{sec:gps}

\begin{figure}
\begin{center}
\includegraphics[width=0.95\linewidth]{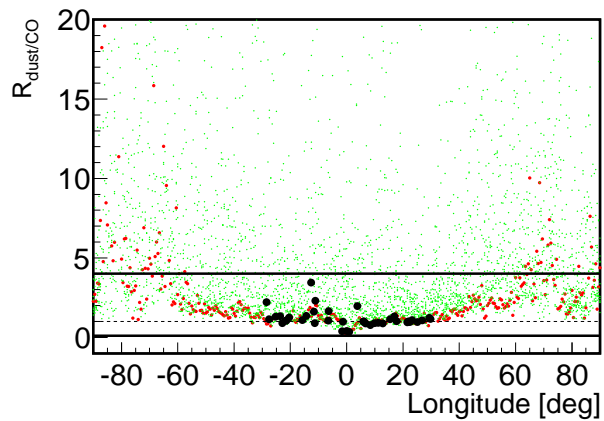}
\caption{Zoom on the VHE sources with Galactic longitude $|l|<30^\circ$: all of the GPS sources are within the allowed range of parameters. }
 \label{fig:gpsratiosigma}
\end{center}
\end{figure}

We select all sources in TeVCat  with ($|l|<30^\circ$ and $|b|<2^\circ$, the Galactic plane survey (GPS) sample). We have computed 
$R_\mathrm{dust/CO}$ for all the projected locations of this sample and all values are in the allowed range. Therefore, no excess or dark gas is needed in these environments. The mass estimate from the molecular component and from the dust tracer are in agreement, if the conversion factor $X_\mathrm{CO}$ is allowed to vary, as shown in Fig. \ref{fig:gpsratiosigma}.
RXJ1713.7--3946 is one of the sources close to the upper border. This is indeed a common example of an environment in which the HI atomic component needs to be taken into account when modeling the VHE emission through hadronic processes \citep[see ][]{fukuirxj}.

\subsubsection{Gould belt clouds}

We have calculated the $R_\mathrm{dust/CO}$ quantity for the projected location of the Gould clouds. We use the same procedure outlined in Section \ref{sec:estimation}, but considering the full extent of the cloud in the survey border of Table 1 in \citet{newdame_co}.
We use again the $\wco$ information from the Planck ``Type2'' CO all-sky maps and Planck dust maps.
Table \ref{tbl:clouds} summarizes the mass estimates for these objects. The latter are calculated through Eqs. \ref{eq:massco} and \ref{eq:massdust}, using the distances to the clouds given in \citet{dame_co} and $X_\mathrm{CO} = 1.8 \times 10^{20} \textrm{cm}^{-2} (\textrm{K km/s})^{-1}$ \citep[as used in][]{newdame_co} and $\sigma_{high}(353\mathrm{GHz})=4.1 \times 10^{-26}$ cm$^2$ H$^{-1}$ \citep[appropriate for a large column density typical of molecular clouds, as in][]{roy_orion_dust}.
When considering the $R_\mathrm{dust/CO}$ value for all the clouds in Table \ref{tbl:clouds}, no excess gas needs to be invoked in these environments to reconcile the mass estimate from CO and dust, with the exception of the Ophiucus cloud.
The existence of \textit{``CO-faint''} gas can be probed also through HE \grs (HE; E$\gtrsim$100 MeV), from the assumption that the HE flux is due only to the interaction of CR and the target mass. The case of molecular cloud emission in HE \gr and the relation to their mass content is explored in \citet{ruizhi_fermi_clouds}, where it is also found that no excess or dark gas is needed in these high density regions\footnote{Some of the mass estimates given here are different by a factor of $\sim2$ from those calculated in \citet{ruizhi_fermi_clouds}. This is due to slightly different integration regions used, with the largest exception being the Ophiucus cloud that we integrate in its entirety (a region of $28^\circ \times 14^\circ$).}.

\begin{table*}[!htb]
\centering
\caption{Characteristics of the molecular cloud sample.}
\begin{tabular}[t]{ccccccc}
\hline
Number & Name    &     glon      &    glat   &  M$_\mathrm{CO} $  &  M$_\mathrm{dust}$ & $R_\mathrm{dust/CO}$\\ 
       &         &     (deg)     &    (deg)  &   $10^4(\msol)$       &   $10^4(\msol)$      &      \{0.1 ... 4\}    \\ 
\hline
1&Ophiucus&356.00& 18.00& 1.07& 3.06&4.32\\
2&Hercules& 45.00&  9.00& 1.16& 1.84&2.39\\
3&Orion B&205.00&-14.00& 6.15& 8.16&2.00\\
4&Orion A&213.00&-20.00& 7.00& 7.60&1.63\\
5&Mon R2&214.00&-12.00& 8.53&16.25&2.87\\
6&R CrA&  0.50&-19.00& 0.12& 0.10&1.31\\
7&Chamaeleon&300.00&-16.00& 0.80& 1.09&2.05\\
8&Aquila& 26.00&  7.00& 4.79& 5.34&1.68\\
9&Perseus&158.00&-20.00& 2.87& 2.94&1.54\\
10&Taurus&170.00&-16.00& 2.58& 4.45&2.60\\
\hline
\end{tabular}
\label{tbl:clouds}
\end{table*}
\section{Conclusions}

A large part of the detected Galactic VHE sources is still classified as unidentified, due to the lack of suitable multiwavelength counterparts. We have investigated here the possibility that these unidentified sources are located in peculiar regions of the Galaxy, where the classical estimates of the material content is not complete. We therefore analyzed the possible existence of \textit{``CO-faint''} gas in the location of TeV unidentified sources. The presence of \textit{``CO-faint''} gas can be constrained with a combination of three tracers: intensity of the $^\textrm{12}$CO(1$\rightarrow$0) line to trace the molecular material \citep[e.g.,][]{newdame_co}, dust emission to trace the total hydrogen content \citep[e.g.,][]{planckdust},  and HI hyperfine transition at 21 cm to trace for atomic hydrogen \citep[e.g.,][]{lab_hi}.

There are two main difficulties in these estimations. On one hand, the dust emission is projected in the sky, integrating along the line of sight. As also the distance of the UNID sources is unknown, we would anyway not be able to use information on the distribution along the line of sight. On the other hand, every mass estimate relies on conversion factors that are usually derived with a cross-calibration of the classical tracers. In order to alleviate the influence of the conversion factor on our conclusions, we defined quantities that depend only on a combination of such parameters in Eq. \ref{eq:ratiounsig} and Eq. \ref{eq:threetracers}. We then explored the allowed parameter space of these quantities constraining it with the maximum/minimum values for the conversion factors given in the literature. 

We find for most of the UNIDs, GPS  sources in the inner Galaxy, or molecular cloud complexes, the mass estimate from the molecular component can be reconciled with the mass estimate from the dust component. In the majority of the cases, this requires a combination of a large emissivity parameter for the dust, $\tau_\mathrm{353 GHz}$, and a large $X_\mathrm{CO}$ factor.
For very few UNID sources, located in the outer Galaxy and in particular in the Cygnus region, a variation of conversion factors is not enough and there is the need to precisely consider also the atomic component. This is especially the case for the majority of the locations outside the inner Galaxy (see discussion in Section \ref{sec:define_and_gal}) where the molecular component is clearly not dominant.  We have however shown that no additional gas component is needed to reconcile mass estimation from the three different tracer, when the uncertainty of the conversion factor $X_\mathrm{CO}$, $X_\mathrm{HI}$ and $\sigma(353\mathrm{GHz})$ is considered.

Nonetheless, all possible gas components need to be taken into account when studying the expected contribution to the VHE emission. The precise knowledge of the conversion factor of a region is paramount for precise hadronic emission modeling of VHE source. A larger target mass would relax the requiment for the power of the source of accelerated particles for the same VHE flux. Therefore we have searched here for the possibility of missing mass from the classical mass estimates in the special direction of VHE sources. It is however to be noted that here we can only assign upper limits on the mass target for VHE emission as the material content information are integrated along the line of sight.
High material content and high dust emission is also expected in regions of high stellar activity, where known VHE emitter are probable to appear, like PWN, SNR, binaries or regions of massive-star formation.

The same reasoning would hold in an exploration of sources detected through Fermi-LAT data at HE and without a sure counterpart \citep{2fgl}. However, the source finding algorithm for Fermi-LAT relies on an assumption of the distribution and content of the surrounding gas in order to model the diffuse HE component. The diffuse emission component model includes also a residual intensity map of unmodeled emission (i.e., Galactic diffuse emission not traced by the gas and the Inverse Compton model from GALPROP\footnote{http://galprop.stanford.edu/index.php}). Therefore any contribution in HE \gr from a possible additional untraced gas component would be included in the diffuse component and not in the source flux. 

We have used here coarse maps with a $0.5^\circ$ pixelization, being this the resolution of the LAB HI survey. This was sufficient for the puropose of our study, as it is a similar angular size to the extension of the VHE sources investigated here. However, for a fine hadronic modeling of the VHE emission, a more detailed morphological mapping of the gas content would be required. The forthcoming GASKAP survey of the HI content in the Galactic plane will be extremely useful thanks to its superior sensitivity and especially angular resolution \citep[30'', see ][]{gaskap}. It will provide also distance information necessary for hadronic modeling.
\section*{Acknowledgment}
We acknowledge support from the 
the grants AYA2012-39303 and SGR2014-1073. We thank the anonymous referee for comments that helped improving the paper.


\section*{References}
\begin{thebibliography}{99}

\bibitem[{Abergel et al.}(2011)]{ismplanck}
Abergel, A., {Ade}, P. A. R., Aghanim, N. 2011 A\&A, 536, A21

\bibitem[{Abramowski et al.}(2011)]{1356_hesspaper}
Abramowski, A., Acero, F., Aharonian, F. (\hess Collaboration) 2011, A\&A, 533A, 103H

\bibitem[{Ackermann et al.}(2012)]{cygnusfermi}
Ackermann, M., Ajello, M., Allafort, A. et al. 2012, A\&A, 538, 71

\bibitem[{Ade et al.}(2014)]{planckdust}
{Ade}, P. A. R., Aghanim, N., Alves, M. I. R. et al. 2014, A\&A, 564A, 45P

\bibitem[{Ade et al.}(2013)]{planckco}
{Ade}, P. A. R., Aghanim, N., Alves, M. I. R. et al. 2013, arXiv:1303.5073 

\bibitem[{Ade et al.}(2011)]{planckdark}
{Ade}, P. A. R., Aghanim, N., Arnaud, M. et al. 2011, A\&A 536, A19 

\bibitem[{Aharonian et al.}(2006)]{surveyhess}
Aharonian, F., Akhperjanian, A. G., Bazer-Bachi, A. R., et al. 2006, ApJ, 636, 777

\bibitem[{Bianchi}(2013)]{bianchi_beta}
Bianchi, S. 2013, A\&A, 552A, 89B

\bibitem[{Bolatto et al.}(2013)]{xcoreview}
Bolatto, A. D., Wolfire, M. \& Leroy, A. K. 2013, ARA\&A,51, 207B

\bibitem[{Carrigan et al.}(2013)]{hgps_icrc13}
{Carrigan}, S., Brun, F., Chaves, R. C. G., et al. 2013, arXiv:1307.4868

\bibitem[{Coupeaud et al.}(2011)]{macstudy}
{Coupeaud}, A. 2011, A\&A, 535A, 124C

\bibitem[{Dame et al.}(1987)]{dame_co}
{Dame}, T. M. et al. 1987, ApJ, 322, 706

\bibitem[{Dame et al.}(2001)]{newdame_co}
{Dame}, T. M., Hartmann, Dap, Thaddeus, P. 2001, ApJ, 547, 792D

\bibitem[{Dickey et al.}(2013)]{gaskap}
{Dickey}, J. M. et al.2013 PASA, 30, 3D

\bibitem[{Dunne et al.}(2009)]{casa_dust}
Dunne, L., Maddox, S. J., Ivison, R. J., et al. 2009, MNRAS, 394, 1307D

\bibitem[{Fukui et al.}(2012)]{fukuirxj}
Fukui, Y. et al 2012, ApJ, 746, 82

\bibitem[{Fukui et al.}(2014)]{fukuithick}
Fukui, Y. et al. 2014 arXiv:1401.7398

\bibitem[{Grenier et al.}(2005)]{grenierdark}
Grenier, I. A.. Casandjian, J., Terrier, R. 2005, Science, 307, 1292

\bibitem[{Kalberla et al.}(2005)]{lab_hi}
Kalberla, P. M. W., Burton, W. B., Hartmann, D., et al. 2005, A\&A, 440, 775

\bibitem[{Nolan et al.}(2012)]{2fgl}
Nolan, P. L. et al. 2012, ApJS, 199, 31N


\bibitem[{Roy et al.}(2013)]{roy_orion_dust}
Roy A. et al. 2013, ApJ, 763, 55

\bibitem[{Yang et al.}(2014)]{ruizhi_fermi_clouds}
Yang, R., de O\~{n}a Wilhelmi, E., \& Aharonian, F. 2014, A\&A, 566A, 142Y

\end{thebibliography}
\end{document}